\documentclass[twocolumn]{aa} 

\usepackage{tikz}
\usepackage{graphicx}
\usepackage{txfonts}
\usepackage[switch]{lineno} 
\usepackage{hyperref}
\hypersetup{
	colorlinks=true,
	linkcolor=blue,
	filecolor=magenta,      
	urlcolor=cyan,
	citecolor=blue,        
}
\begin{document} 
   \title{CO$^+$ first-negative band emission: A tracer for CO in the 
   	Martian upper atmosphere}
   \subtitle{}
   \author{Susarla Raghuram\thanks{Corresponding 
   author \email{raghuramsusarla@gmail.com}}
         and Anil Bhardwaj}
   \institute{Physical Research Laboratory, Ahmedabad, 380009, 
   	India.}

\titlerunning{CO$^+$ first-negative band emission in the Martian upper 
atmosphere}
\authorrunning{Raghuram and Bhardwaj}

 
  \abstract
   {Recently, Imaging Ultraviolet Spectrograph (IUVS) on-board 
   Mars Atmosphere and Volatile EvolutioN (MAVEN) satellite observed
   CO$^+$ first-negative band limb emission  in the Martian 
   upper atmosphere.}
   {We aim to explore the photochemical processes in the Martian upper 
   	atmosphere, which drives this band emission.}
   {A photochemical model is developed to study	the excitation processes 
   of CO$^+$ 
   	first-negative band emission (B$^2\Sigma^+ \rightarrow$ X$^2\Sigma^+$) 
   	in the 
   	upper atmosphere of Mars.  The number density profiles of CO$_2$ and 
   	CO from two
   	different models, viz., Mars Climate Database (MCD) and
   	Mars Global Ionosphere-Thermosphere (MGIT) are used to determining
   	the limb 
   	intensity of this band emission.}
   {On increasing the 
   	CO density by a factor of {4} and 8 in MCD and MGIT models, 
   	respectively, the 
   	modelled	CO$^+$ first-negative band limb 
   	intensity profile is found to be consistent with IUVS/MAVEN 
   	observation.  In this case, the intensity of this band emission
   	is significantly determined by	the ionization  of CO by solar photons 
   	and photoelectrons, and the	role of dissociative ionization of CO$_2$ 
   	is negligible.}
   {Since CO is the major source of {the CO$^+$(B$^2\Sigma^+$)}, we 
   suggest that the observed CO$^+$ first-negative band emission 
   	intensity can be used to retrieve {the} CO  
   	density in the Martian upper atmosphere for the altitudes above 150 
   	km.}

   \keywords{Molecular processes --
                Planets and satellites: atmospheres -- Planets and 
                satellites: composition }
   \maketitle

\section{Introduction}
Aeronomical emissions of various excited species have been used as 
potential tools to study the composition of the Martian upper atmosphere 
\citep{Barth71, Strickland72, Bertaux05, Leblanc06, Cox10, Simon09, 
  Stiepen15, Evans15, Jain15, Stevens19, Deighan18,
	   Ritter19,Gkouvelis20,Jain20,Qin20}. Besides the emissions of  several 
	   atomic 
	   and molecular
   species, the emissions from the ionic species, such as CO$_2^+$ 
(Fox-Duffendack-Barker band and ultraviolet doublet), N$_2^+$ 
(Meinel and first-negative bands), and CO$^+$ (first-negative and 
Comet tail bands), are also important emission features in the 
ultraviolet spectra. The band heads of CO$^+$ first-negative system
occur nearly at the same wavelengths as that of CO Cameron band emission, 
which makes them challenging to observe in the 
Martian ultraviolet spectra. \cite{Barth71} have reported the 
observation of CO$^+$ first-negative band emission 
(B$^1\Sigma$ $\rightarrow$ X$^2\Sigma^+$) in the Mars during 
Mariner 6 and 7 experiments. However, the detection of this 
band emission in the Mariner 6 and 7 ultraviolet spectra is not
{straightforward}. After subtracting the observed {Martian} 
ultraviolet 
spectra with the identified CO Cameron bands, the residual spectral 
features are ascribed to CO$^+$ first-negative band emissions. 
By constructing the synthetic spectra, \cite{Conway81} has noticed that 
this band emission was absent in the Mariner 9 observation. Recently, 
by analysing the IUVS/MAVEN observations during 
6--8 April 2016,  \cite{Stevens19} reported the first-ever 
CO$^+$ first-negative band emission in the upper atmosphere of Mars. 


The photochemical processes of the 
principal emissions of Martian upper atmosphere have been studied in 
several works by accounting for the various formation and loss mechanisms 
of different excited atmospheric species, \citep{Fox79, Mantas79, 
Conway81, Shematovich08, Simon09, Gronoff12a, Gronoff12b, Jain12, Jain11, 
Gkouvelis18, Gerard19, Ritter19}. Photoionization of 
CO, photoelectron impact ionization of 
CO, photodissociative ionization of CO$_2$, electron impact 
dissociative ionization of CO$_2$, and resonance fluorescence of CO$^+$ 
are the  excitation sources of CO$^+$ in B$^2\Sigma^+$ state. By 
assuming 
different photon and 
electron impact cross sections of CO and CO$_2$, \cite{Conway81} 
estimated that 
photodissociative ionization of CO$_2$ is the major source of this band 
emission in the Martian upper 
atmosphere. In light of the 
recent IUVS/MAVEN observation of CO$^+$ 
first-negative 
band emissions, we have developed a photochemical model to study 
the emission processes of CO$^+$ first-negative band in the 
Martian upper atmosphere. The main aim of the present work is to study the 
contribution  of different excitation processes which {produce} CO$^+$ 
first-negative band emission during the IUVS/MAVEN observation.
By considering different atmospheric neutral density profiles in the 
model, we {determine} the volume emission rates of 
CO$^+$(B$^2\Sigma^+$)  
and the 
limb 
intensity profiles of CO$^+$ first-negative band emission for various 
excitation mechanisms. We also compare our modelled limb intensity 
profiles 
with the IUVS/MAVEN observation.

%

\section{Model inputs and calculations}
\label{sec:model}
\begin{figure}
	\centering
	{\includegraphics[width=\columnwidth]{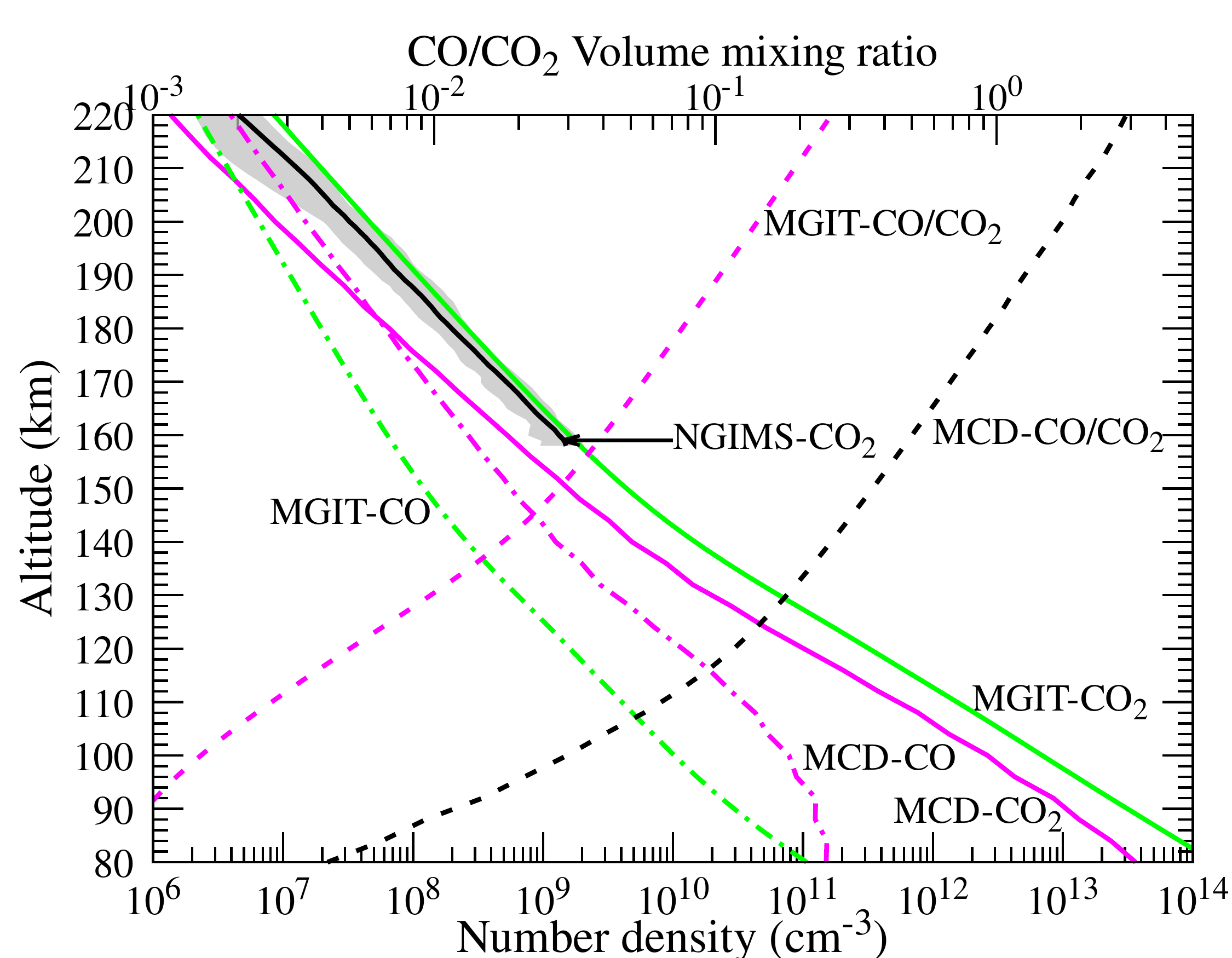}}	
	\caption{Number density profiles of CO$_2$ and CO from MCD and MGIT 
	model. Solid and dash-dotted curves 
	represent CO$_2$ and CO density profiles, respectively. The gray 
	shaded area represents the variation in the NGIMS/MAVEN measured 
	CO$_2$ density during 6--8 April 2016 for the 
	orbits 	\#2953--2962. The solid black curve in the shaded area 
	represents the averaged value of CO$_2$ from the NGIMS/MAVEN 
	measurements. The calculated CO/CO$_2$ {volume mixing} ratio 
	profiles are 
	plotted with 
	dashed curves with scale on top x-axis.}
	\label{fig:numb_neu}
\end{figure}

During the IUVS/MAVEN observation period, i.e. during 6--8 April 2016, 
{the}
Neutral Gas and Ion Mass 
Spectrometer (NGIMS) onboard  MAVEN mission is also measuring the 
neutral density profiles of major species. But the NGIMS measurements of 
the major neutral densities are limited up to the MAVEN periapsis  
altitude of 
about 160 km. Since the peak intensity of CO$^+$ first-negative emission 
{is limited down to} the altitude of 160 km, we used the primary neutral 
densities 
(CO$_2$, CO, 
N$_2$, O$_2$ and O) from {the}
Mars Climate Database 
(MCD) \citep[version 5.3,][]{Gonzalez15} and  Mars 
Global Ionosphere-Thermosphere (MGIT)  \citep{Bougher15} models for 
the IUVS/MAVEN observation	conditions. To compare the MCD and MGIT 
modelled CO$_2$ 
neutral density with in-situ measurements, we analysed the 
NGIMS/MAVEN  level 2 (L2), version 8, revision 1 
data  during 6--8 April 2016 for the 
altitudes 
above 160 km \href{https://pds-atmospheres.nmsu.edu}
{(https://pds-atmospheres.nmsu.edu)}. More details 
of the L2 data product are 
available in \cite{Benna18}. The NGIMS/MAVEN measured neutral 
number densities are interpolated over a uniform grid of 1 km from the 
periapsis altitude, i.e., 155 km to 230 km for all the orbital profiles 
observed during 6--8 April 2016. Figure~\ref{fig:numb_neu} shows the 
CO$_2$ and CO neutral density profiles obtained from the two models. 
In this figure,  MGIT and MCD 
modelled CO$_2$ densities are also compared with the NGIMS/MAVEN 
in-situ measurements. 

\begin{figure}[h!]
	\centering
	\includegraphics[width=1.02\columnwidth]{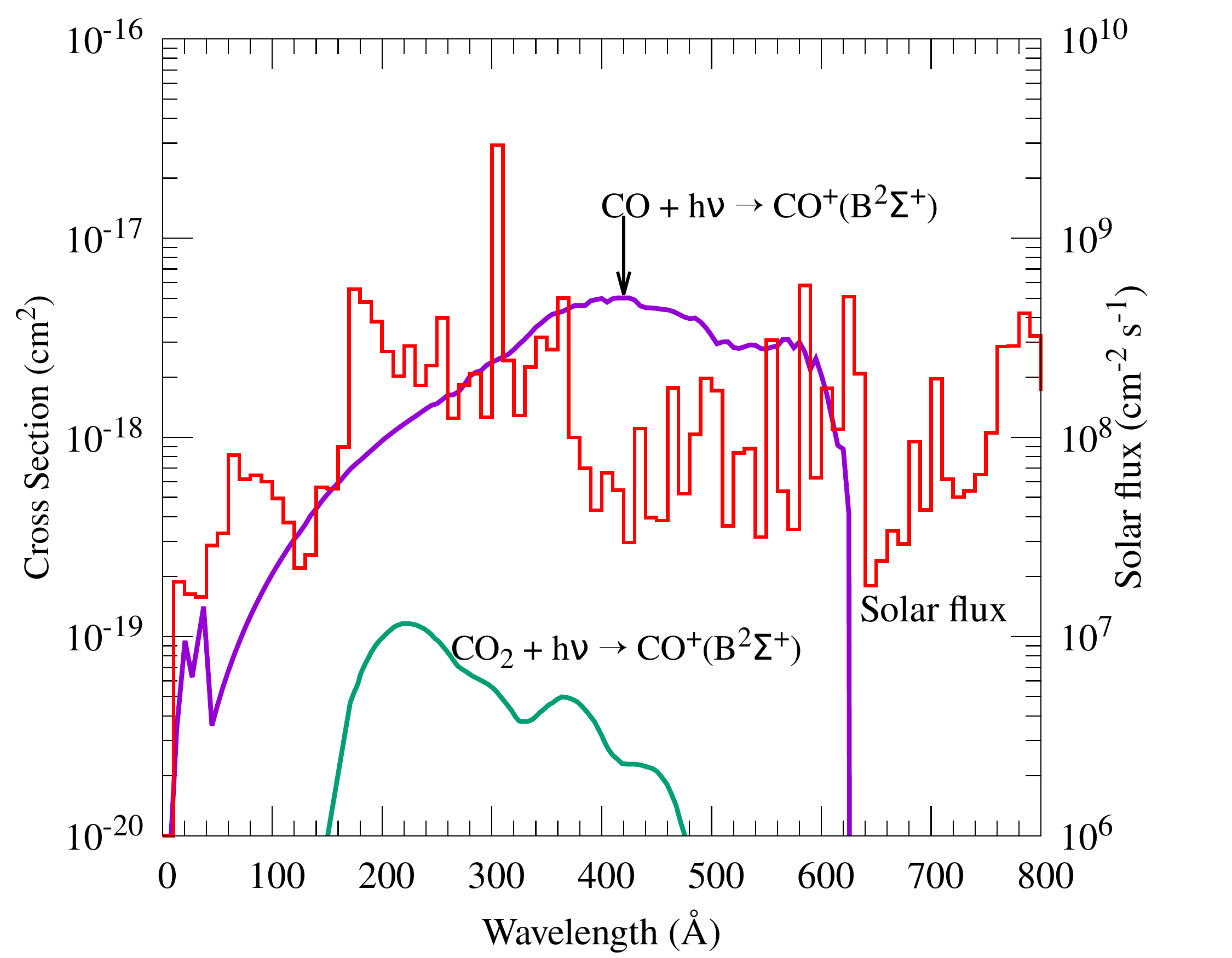}	
	\caption{Photoionization cross sections of 
		CO$_2$	
		and CO producing CO$^+$(B$^2\Sigma^+$), which are taken from 
		\cite{Wu86} and \cite{Plummer77}, respectively. The cross
		section of \cite{Wu86} is linearly extrapolated for the wavelength 
		below 180 \AA.
		The daily-averaged solar flux for the wavelengths below 800 \AA,
		which is obtained	from the measurements of Solar Extreme 
		Ultraviolet 
		Monitor instrument 	onboard MAVEN mission on 7 April 2016 for the 
		orbit 
		\#2960 \citep{Eparvier15}, is plotted {with scale} on the right 
		y-axis.}
	\label{fig:xs}
\end{figure}

\begin{figure}[h!]
	\centering
	\includegraphics[width=1.02\columnwidth]{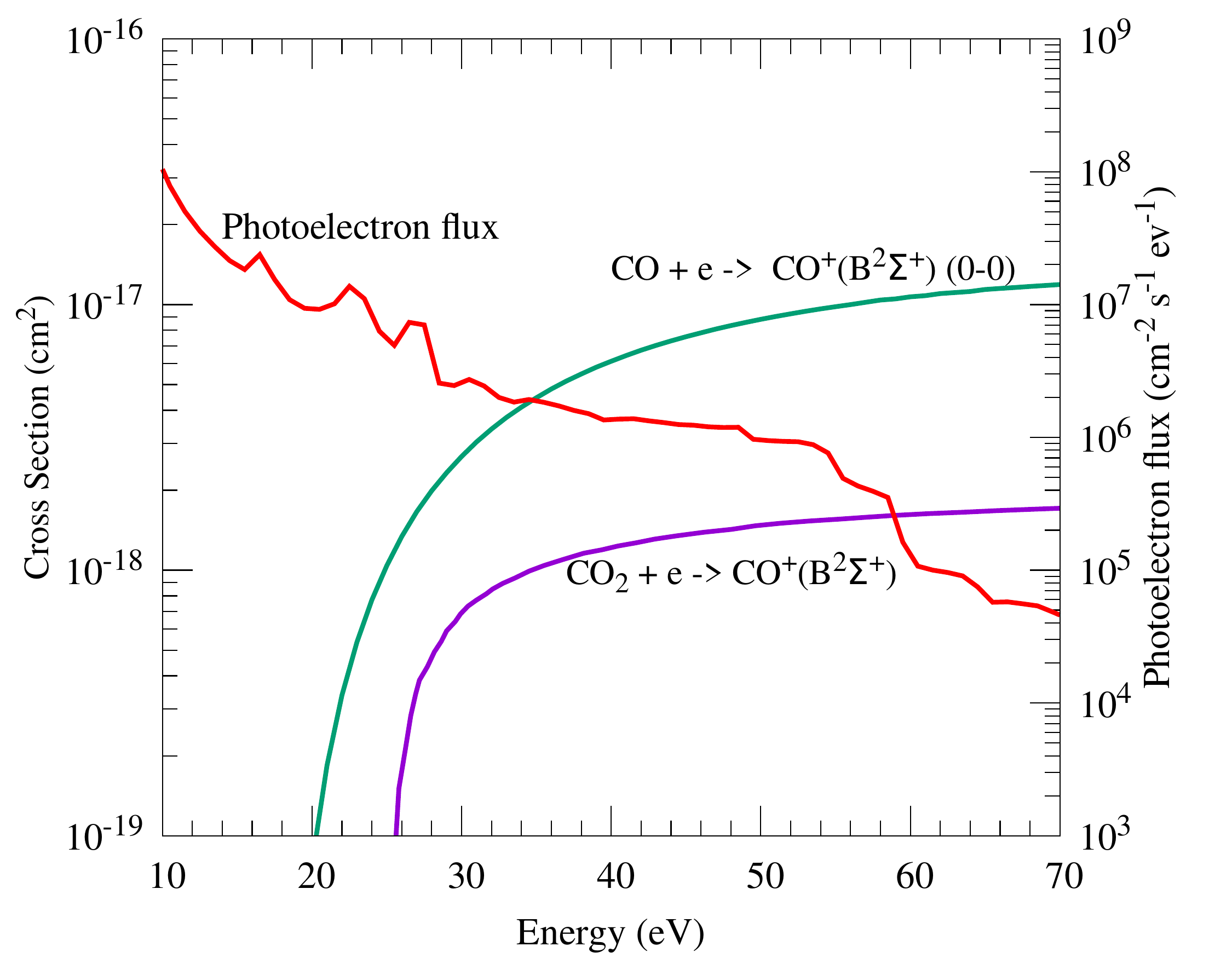}	
	\caption{Electron impact ionization cross 
		sections of 
		CO$_2$ and CO producing CO$^+$(B$^2\Sigma^+$) which are taken from 
		\cite{Ajello71-co2} and \cite{Arqueros81}, respectively.
		Modelled steady state suprathermal 
		electron flux (red solid curve) at an altitude of 130 km using 
		MCD neutral density profile and for  
		SZA of 75$^\circ$ is plotted with a 
		scale on the right 	y-axis.}
	\label{fig:exs}
\end{figure}

Solar radiation flux, which is the primary energy source of Martian upper 
atmosphere, is taken from Solar Extreme 
Ultraviolet 
Monitor (EUVM) instrument onboard 
MAVEN mission  measured daily-average value in the 
wavelength region 
5--1900 \AA\ on 7 April 2016 for 
orbit \#2960  {\citep{Eparvier15,Thiemann17}}.  A part of this solar flux 
spectrum is plotted 
in  
Figure \ref{fig:xs}.   The 
photoabsorption and photoionization cross sections of  the primary gases 
are taken from \cite{Huebner92} 
\href{https://phidrates.space.swri.edu}
{(https://phidrates.space.swri.edu)}. 
The 
branching ratios 
for 
different excited states of CO$_2^+$, CO$^+$, N$_2^+$, O$^+$, and O$_2^+$ 
ions are
taken from \cite{Avakyan98}. 
The photoionization and dissociative photoionization excitation cross 
sections of CO and CO$_2$ producing CO$^+$(B$^2\Sigma^+$) are taken from 
\cite{Plummer77} and \cite{Wu86}, respectively. These cross sections are 
also plotted in  Figure~\ref{fig:xs}. The electron impact cross 
sections of CO$_2$ and CO 
producing CO$^+$(B$^2\Sigma^+$) are obtained from \cite{Ajello71-co2} 
and 
\cite{Arqueros81}, respectively. These cross sections are presented in 
 Figure~\ref{fig:exs}. The branching ratio for 
CO$^+$(B$^2\Sigma^+$) that 
leads to (0-0)  electronic transition is taken as 0.305 
\citep{Arqueros82}. 
Under photochemical equilibrium condition, the densities of major 
(O$_2^+$, 
NO$^+$) and minor  (CO$_2^+$, CO$^+$, N$_2^+$, O$^+$, 
and C$^+$) ions are also calculated by accounting for the important 
production 
and loss mechanisms. {Resonance} fluorescence of CO$^+$ 
producing first-negative band emission is incorporated in the model by 
multiplying the modelled CO$^+$ ion density with an emission rate 
(g-factor) of 
5.2 $\times 10^{-5}$ s$^{-1}$ \citep{Barth69}. 
\cite{Scholl98} experimentally 
determined the radiative lifetime of B$^2\Sigma^+$ excited state as about 
55 
ns. Due to the short radiative lifetime, the collisional quenching of this 
excited state can be neglected.
A fractional 
population of this excited 
state can decay to the ground state via A$^2\Pi_u$ which leads to 
comet-tail 
band emission.
\cite{Lawrence65} has determined that about 10\% of 
CO$^+$(B$^2\Sigma^+$) ions decay to A$^2\Pi_u$ state and the rest 
directly decays to the ground state. {Hence, 
we consider 90\% of modelled volume emission rate leads to CO$^+$ 
first-negative band 
emission.}

{The modelled suprathermal electron flux in the Martian upper atmosphere 
at an 
altitude of 130 km is presented in Figure~\ref{fig:exs}.
Detail explanation for the calculation of 
steady-state suprathermal electron flux and limb emission intensities in 
the 
Martian upper atmosphere is given in our earlier works 
\citep{Jain11,Jain12}. Here, we briefly describe 
the 
calculation of suprathermal electron flux spectra.
Solar radiation is degraded using 
Beer-Lambert's law in the Martian upper atmosphere and primary 
photoelectron 
production rate spectrum Q(Z, E), at a given altitude Z as a function of 
energy 
E, is calculated using the following equation.}

\begin{equation}
Q(Z, E) = \sum_l n_l(Z) \sum_{j,\lambda} \sigma_l^I(j,\lambda)
I(Z,\lambda) \delta\left(\frac{hc}{\lambda} - E - W_{jl}  \right) 
\end{equation}

\begin{equation}
I(Z,\lambda) = I(\infty, \lambda) exp\left[-sec(\chi) \sum_l 
\sigma_l^A(\lambda) \int_{Z}^{\infty} n(Z') dZ'  \right]
\end{equation}

{Where $\sigma_l^A(\lambda)$ and $\sigma_l^I(j, \lambda)$ are the 
total photoabsorption and  photoionization cross
sections of $j^{th}$ ion of the $l^{th}$ atmospheric species
at wavelength $\lambda$,  respectively. $I(\infty, \lambda)$ is the 
unattenuated solar flux at wavelength $\lambda$ at the top of the 
atmosphere. $n_l(Z)$ is the number density of $l^{th}$ species
at altitude $Z$. $\chi$ is the solar zenith angle (SZA); 
$\delta$(hc/$\lambda$ - E - $W_{jl}$) is the delta function, in which 
hc/$\lambda$ is the energy of incident photon, $W_{jl}$ is the 
ionization threshold of $j^{th}$ ionic state of the $l^{th}$ species,
and $E$ is the energy of ejected electron. 
We have  calculated the steady state photoelectron flux $\phi(Z,E)$ at 
altitude $Z$ and at energy $E$ using Analytical Yield Spectrum (AYS)
method using the following equation.} 

\begin{equation}
\phi(Z,E) = \int_{W_{kl}}^{100} \frac{Q(Z,E) U(E,E_0)}{\sum_l n_l(Z) 
	\sigma_{lT}(E)} dE_0 
\end{equation}

{Where, $\sigma_{lT}(E)$ is total inelastic cross section of $l^{th}$ 
species, and $U(E, E_0)$ is the two-dimensional AYS which contains 
the non-spatial electron degradation information.
More details of AYS method can be found in our earlier work 
\citep{Singhal84, Bhardwaj90, Bhardwaj96, Bhardwaj99a, Jain11, Jain12, 
	Jain13,Raghuram12,Raghuram13}. }

{We have determined the volume emission rate profiles of 
CO$^+$(B$^2\Sigma^+$) for different electron 
and photon induced processes using the  modelled photon and 
suprathermal electron fluxes and the corresponding cross sections.}
These modelled volume emission rate profiles  are 
integrated along the IUVS/MAVEN tangential line of sight and converted 
into 
brightness in Rayleigh (1 Rayleigh = 10$^6$/4$\pi$ photons cm$^{-2}$ 
sec$^{-1}$ 
sr$^{-1}$) using the following equation.

\begin{equation}
I = 2 \times  10^{-6} \int V(r) dr
\end{equation}

Here r is the abscissa along the horizontal line of sight, and 
V(r) is the volume emission rate (photons cm$^{-3}$ s$^{-1}$) at a 
particular emission point r on the tangent. The factor of 2 multiplication 
comes due to the symmetry along the line of sight  concerning the tangent 
point. {This factor is necessary when the  volume 
emission rate is integrated from the tangent point to infinity.} The absorption 
of CO$^+$ 
first-negative band emission by other 
Martian 
species along the line of sight is negligible due to the low 
photoabsorption 
cross section. 

\begin{figure*}
	\centering
		{\includegraphics[width=\columnwidth]{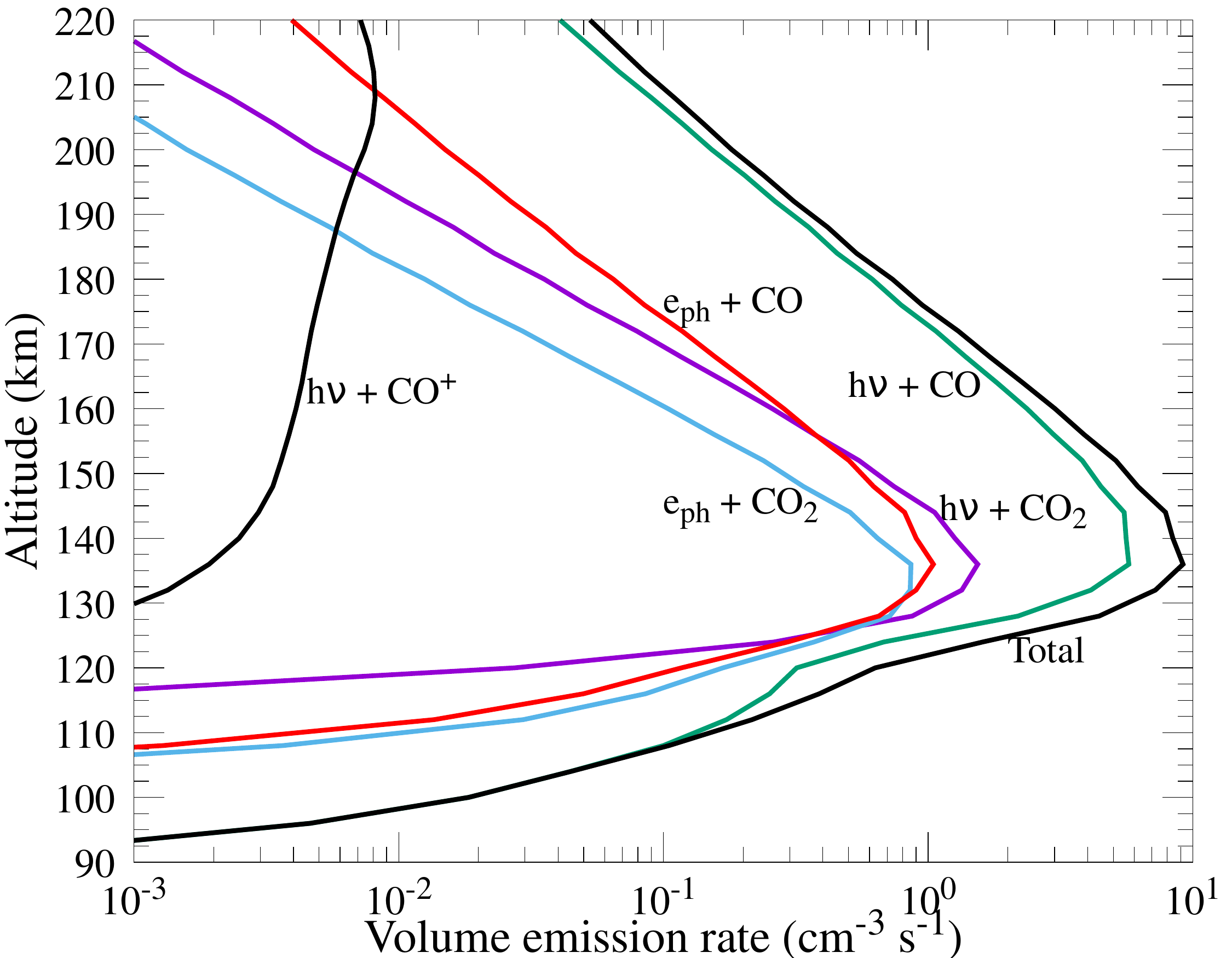}}
	{\includegraphics[width=\columnwidth]{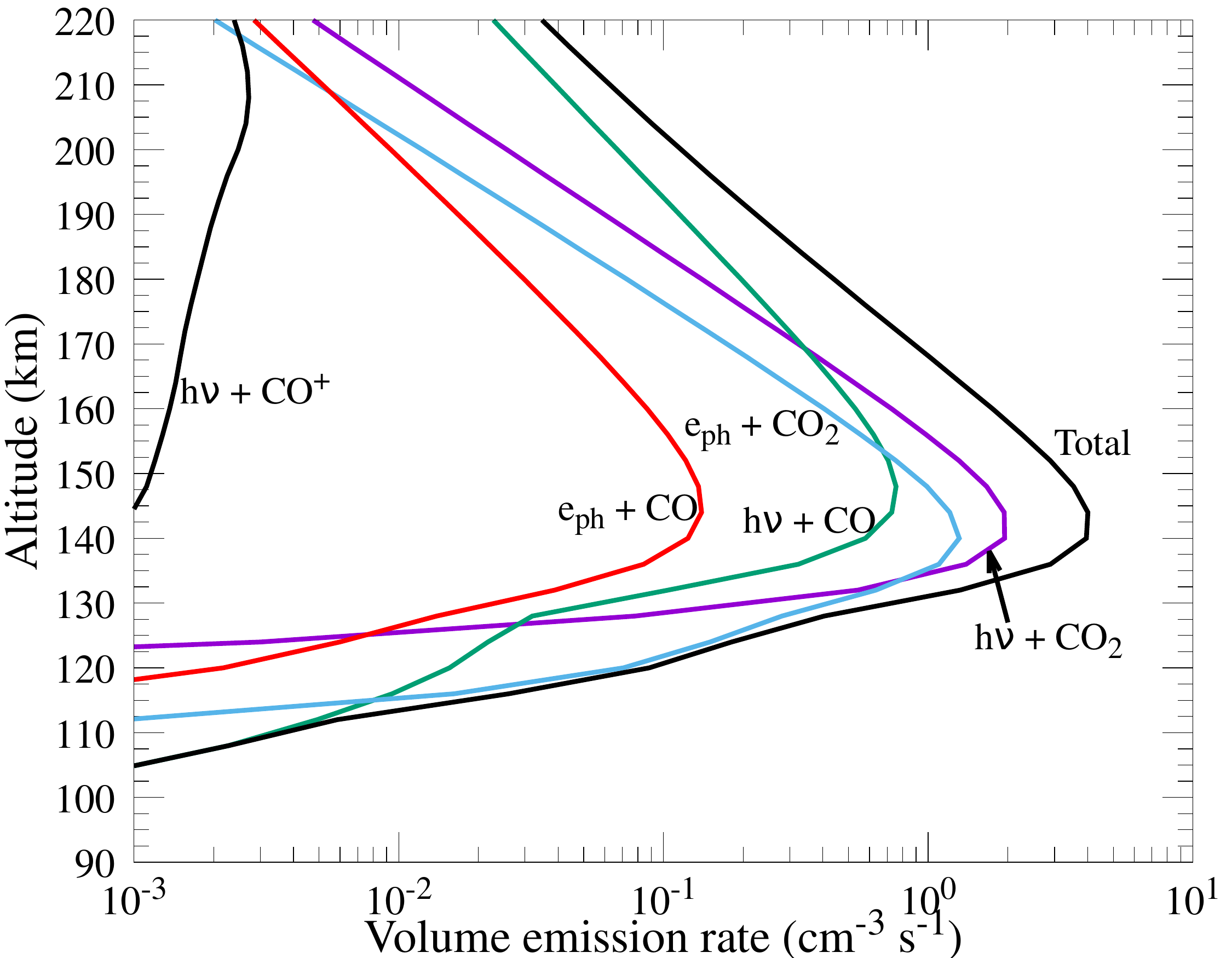}}
		\caption{Calculated volume emission rate profiles of 
		CO$^+$(B$^2\Sigma^+$) in the Martian upper atmosphere via 
		various production mechanisms  using neutral density profiles from 
		MCD  (left panel) MGIT (right panel) models  for SZA 
		75$^\circ$. 
		h$\nu$ and e$_{ph}$ represent solar photon and 	suprathermal 
		electron, respectively.}
	\label{fig:vol_prate}
\end{figure*}

\section{Results and Discussion }
\label{sec:res_dis}

The left and right panels of Figure~\ref{fig:vol_prate}  show the 
modelled volume emission rate profiles of CO$^+$(B$^2\Sigma^+$) via 
different excitation mechanisms using the neutral density profiles from 
MCD and MGIT models, respectively, for the IUVS/MAVEN observational 
conditions at SZA of 75$^0$. These 
calculations show that peak formation rate of CO$^+$(B$^2\Sigma^+$) 
occurs at  an altitude of around 140 km. On using the 
neutral densities from 
MCD model, the total volume emission rate of 
CO$^+$(B$^2\Sigma^+$) is significantly controlled by photoionization of 
CO and the contribution from photon and electron impact dissociative 
ionization processes of CO$_2$ to the total is less than 30\% (see the 
left 
panel of Fig.~\ref{fig:vol_prate}).
When we use the neutral density profiles from MGIT model, about 80\% of 
the total peak volume emission rate of CO$^+$(B$^2\Sigma^+$) is controlled 
by dissociative ionization of CO$_2$, {whereas} photoionization of CO 
is the 
important production source for the altitudes above 170 km (see the right 
panel of 
Fig.~\ref{fig:vol_prate}).

It can be noticed in Figure~\ref{fig:numb_neu} that the CO$_2$ (CO) 
density profile of MGIT  model is higher (lower) by about a factor of 
3 (5) compared to that of MCD model. 
Hence, the differences in neutral densities 
of MCD and MGIT models lead to the photoionization of CO and 
photodissociative ionization of CO$_2$ to be respective major
excitation sources of CO$^+$(B$^2\Sigma^+$)  at an altitude of around 140 
km (see Fig.~\ref{fig:vol_prate}). Moreover, the difference in  photon
and electron impact cross sections of CO and CO$_2$ also plays an 
important role in determining the volume emission rate, which will be 
discussed later.   
The contribution from  solar resonance fluorescence of CO$^+$ is smaller 
by about an order of 
magnitude or more to the total volume emission rate in both the modelled 
profiles.

\begin{figure*}[h!]
	\centering	
	{\includegraphics[width=\columnwidth]{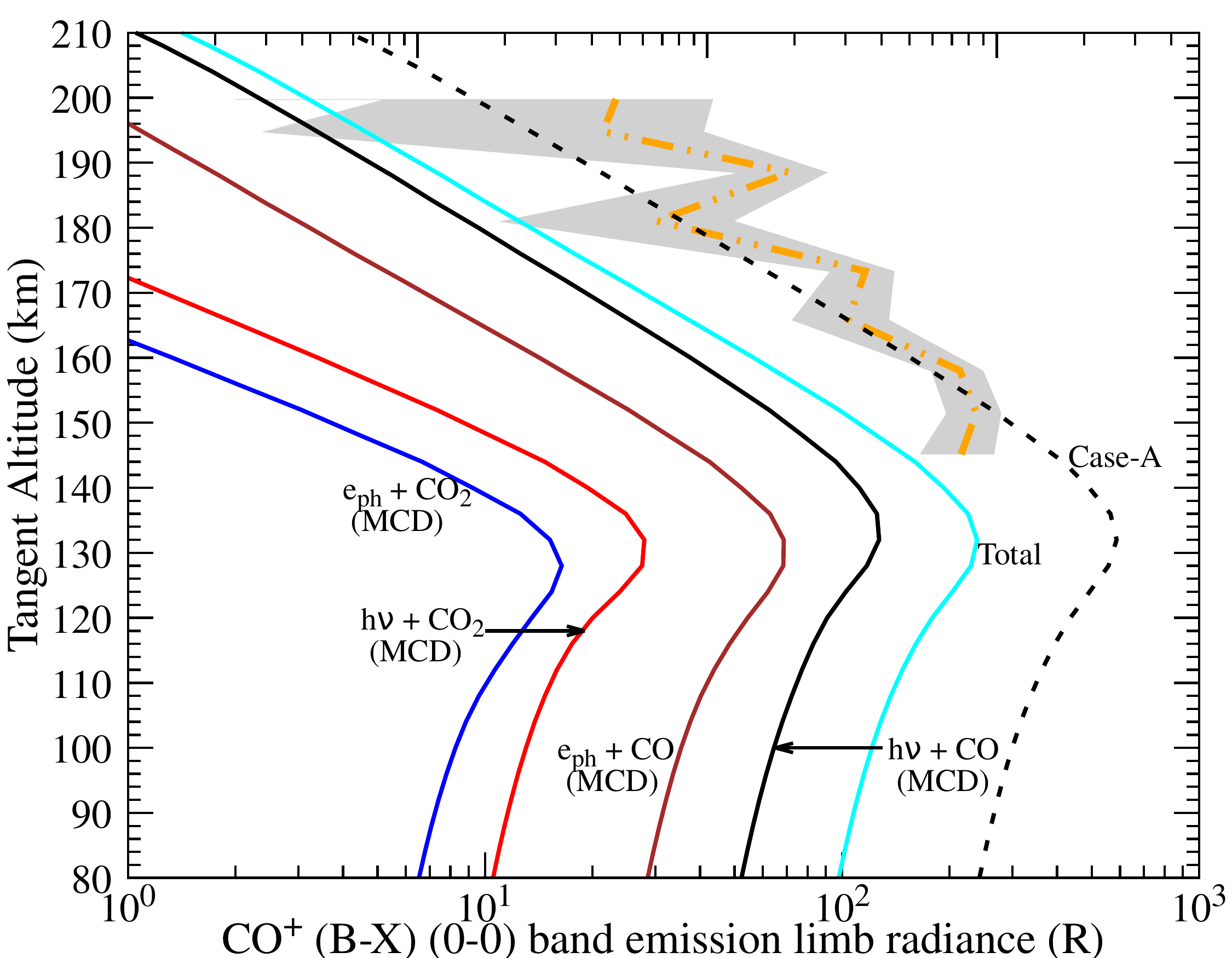}}
    {\includegraphics[width=\columnwidth]{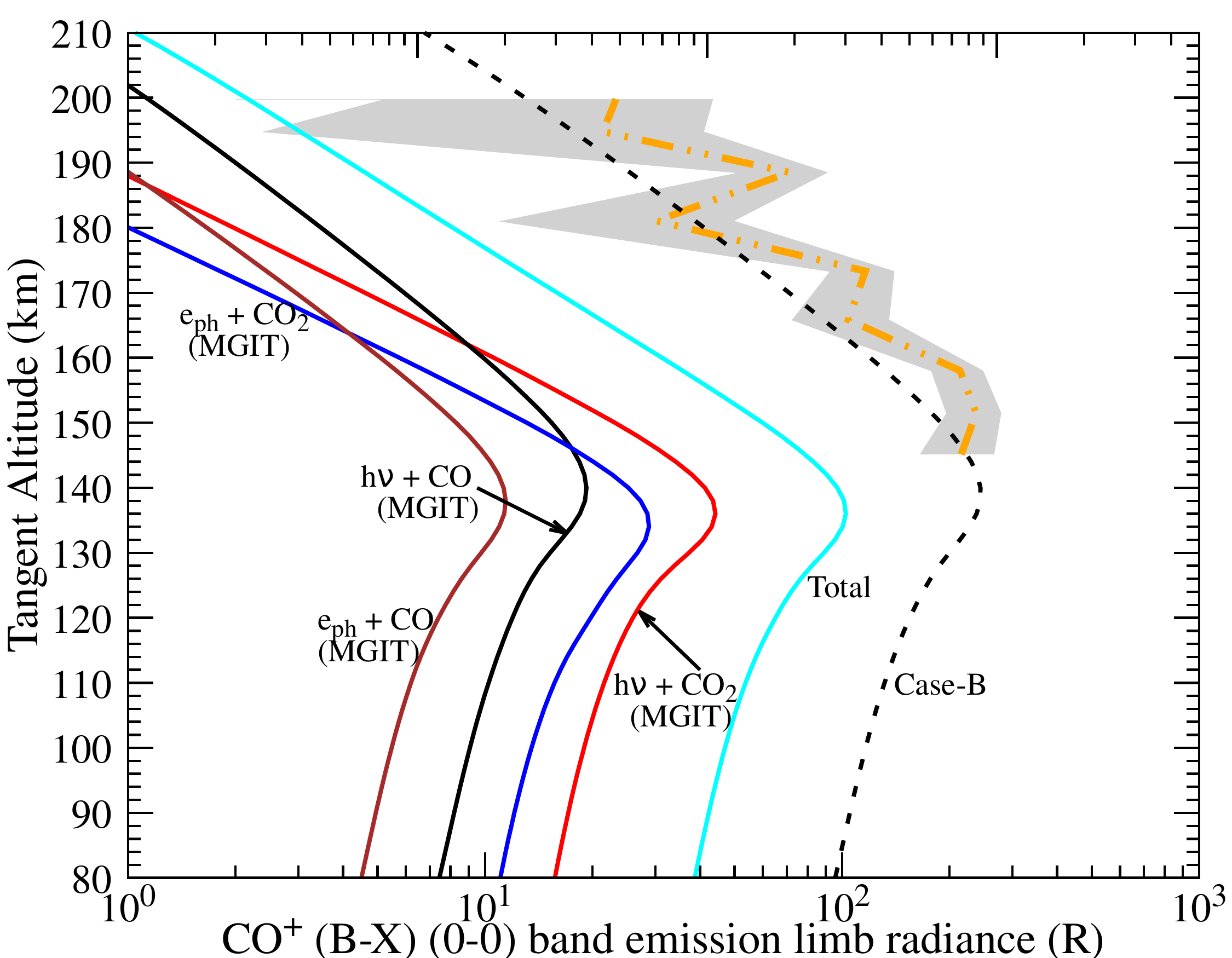}}		
	\caption{Comparison between  modelled CO$^+$ 
		first-negative (0-0)  band emission limb intensity profiles for 
		different excitation mechanisms, using neutral densities from 
		MCD (left panel) and MGIT (right panel)	models, {and IUVS/MAVEN 
		observation}. Dash 
		double-dotted orange curve 
		represents the IUVS/MAVEN observed {average intensity of CO$^+$ 
		first-negative (0-0)  
		band emission} on 7 April 2016 and the gray shaded area 
		is its 1-$\sigma$ uncertainty \citep[{taken 
		from}][]{Stevens19}. 
		Here Case-A and Case-B are the modelled limb intensity 
		profiles for the total ionization of CO 
		(photon + photoelectron) by increasing the CO density in 
		MCD and MGIT models by a factor of 3 and 8, 
		respectively. h$\nu$  and e$_{ph}$ 
        represent solar photon and suprathermal electron, respectively.}
	\label{fig:com_fmin}
\end{figure*}

Figure~\ref{fig:com_fmin} shows a comparison between the modelled limb 
intensities for different excitation mechanisms, which are determined
using neutral densities from MCD (left panel) and MGIT (right panel) 
models, and IUVS/MAVEN observation. 
Using the neutral density profiles from MCD model, the modelled total 
limb intensity profile is consistent with the lower limit of IUVS/MAVEN 
observation only at a few altitudes (see the left panel of 
Fig.~\ref{fig:com_fmin}).
In this case, photon and photoelectron impact ionization processes  of 
 CO  are the 
significant production sources to the total CO$^+$ first-negative band 
emission  
intensity. As shown in Figure~\ref{fig:numb_neu}, the CO$_2$ neutral 
density profile is smaller by a factor of 3 compared to the averaged value 
of NGIMS/MAVEN 
in-situ measured density for the altitudes above 160 km. To 
assess the role of CO$_2$ on the 
modelled limb intensity, we increased its neutral density profile by a 
factor of 3. On increasing the MCD modelled CO$_2$ density profile, the 
calculated total limb intensity profile is not consistent with the 
IUVS/MAVEN observed average intensity. But when we increase the CO neutral 
density profile by a factor of 3 (hereafter case-A), the modelled limb 
intensity profile is 
in agreement with IUVS/MAVEN observation. These calculations suggest that 
CO density plays a significant role compared to that of CO$_2$ to 
explain the IUVS/MAVEN observed CO$^+$ first-negative band emission 
profile.  

On using the neutral densities from MGIT model, the modelled total limb 
intensity is smaller by a factor of 3 or more compared to the  
average intensity profile of IUVS/MAVEN observation (see the right panel 
of 
Fig.~\ref{fig:com_fmin}). In this case, below 
140 km altitude, photon and electron impact 
dissociative ionization processes of CO$_2$ are the significant sources 
of CO$^+$ 
first-negative band emission. Above 160 km radial distance, the 
contribution from CO is important to the total emission intensity. 
Since the CO$_2$
 neutral density of MGIT model is consistent with NGIMS/MAVEN in-situ 
measurements (see Fig.~\ref{fig:numb_neu}), we assess the impact of CO 
density on the modelled limb emission intensity. By increasing 
the MGIT CO neutral density profile by a factor of 8 (hereafter 
Case-B), the modelled limb 
intensity is found to be in agreement with the IUVS/MAVEN observation. 
This calculation also suggests that a higher amount of CO is 
required to explain the IUVS/MAVEN observed emission profile. By 
considering the variability in  CO$_2$ density as measured by 
NGIMS/MAVEN (a factor of 2--5, as shown in Fig.~\ref{fig:numb_neu}), 
no agreement was found between the modelled limb intensity profile for 
the 
dissociative ionization of CO$_2$ and the IUVS/MAVEN observation,  which 
also suggests that CO$_2$ is not a suitable 
candidate to explain the observed CO$^+$ first-negative band emission.
For the above 
all cases, the contribution from solar resonance is very 
small ($<$1 R) hence it is not shown in Figure~\ref{fig:com_fmin}.

We also explored the role of thermosphere temperature, which determines 
the scale {heights of CO$_2$ and CO densities}, during the IUVS/MAVEN 
observation period.  By analysing the 
NGIMS 
and IUVS measurements, \cite{Bougher17} have studied the structure and 
variability of Martian dayside 
thermosphere during MAVEN 
observation period, i.e. October 2014 to May 2016. This analysis shows 
that 
the mean temperature of the Martian upper atmosphere, during the period 
22 
March 2016
to 10 April 2016 for the  orbits \#2873--2974, was around 195 K with a 	
variability of 155--225 K \citep[see Figure 2 of][]{Bougher17}. IUVS/MAVEN 
observed the CO$^+$ first-negative band emission on 7 April 2016 for the 
orbits 
\#2953--2962 \citep{Stevens19}, and during this period the derived mean 
temperature 
of the thermosphere from the MAVEN observations was below 225 K. 
Moreover, 
as 
shown in Figure~\ref{fig:numb_neu}, the variation in  NGIMS/MAVEN 
measured CO$_2$ neutral density is also small (by a factor $\sim$2 to 5). 
Thus, during 
the IUVS/MAVEN observation period, the derived  
lower thermosphere temperature ($<$225 K), lack of substantial 
variability in 
the 
in-situ measured CO$_2$ neutral density,  and the lower contribution of 
dissociation excitation of CO$_2$ (see Fig.~\ref{fig:com_fmin}) also 
confirm that the dissociative 
ionization of CO$_2$ can not be a significant excitation source to explain 
the observed limb intensity profile.

The smaller value of photon and electron impact dissociative excitation 
cross sections of CO$_2$ compared to that of CO is 
also another main reason for CO$_2$  not being a potential source of 
CO$^+$ 
first-negative band emission (see Figs.~\ref{fig:xs} and \ref{fig:exs}). 
{Using the experimentally determined photoionization cross sections and 
onboard measured solar flux, we calculated the unattenuated photoionization 
excitation frequencies 
of CO$_2$ and CO producing CO$^+(B^2\Sigma^+)$ as  5.5 $\times$ 10$^{-10}$ 
and 1 $\times$ 
10$^{-8}$ s$^{-1}$, respectively. Due to the large flux, the 
solar radiation  at the wavelength He II 303 \AA\ significantly 
determines these modelled photoionization excitation frequencies (see 
Fig.~\ref{fig:xs}).
Our calculated photoionization excitation frequencies show that {the 
formation of CO$^+(B^2\Sigma^+)$ via CO channel is  
20 times more {efficient} compared to that of CO$_2$ per molecule}.
{Hence, though the
number density is small, CO significantly controls the CO$^+(B^2\Sigma^+)$ 
formation  and subsequently CO$^+$ first-negative band emission intensity in 
the Martian upper atmosphere for the altitudes above 150 km.} 
Moreover, due 
to the difference in scale 
heights, 
the CO$_2$  neutral density decreases {with altitude} much faster than 
that 
of CO which leads to {a progressive decrease in the contribution from 
CO$_2$}   to 
the total CO$^+$ first-negative band emission {above the 
homopause}	 (see 
Fig.~\ref{fig:com_fmin}).} 

{Besides the role of  number densities, we also studied the 
impact of uncertainties associated with the measurement of 
photon and electron impact ionization cross sections of CO and CO$_2$ and the 
solar 
flux on the modelled limb intensity 
profiles. 
	As mentioned earlier, the photon cross sections of CO and CO$_2$ 
	producing CO$^+$(B$^2\Sigma^+$) are taken from \cite{Plummer77} and 
	\cite{Wu86}, 
	respectively. \cite{Wu86} have not mentioned the uncertainty in their 
	measured photoionization cross section of CO$_2$. Even on increasing this 
	cross 
	section by a factor of 2 no agreement was found between 
	modelled limb intensity profile and the IUVS/MAVEN 
	observation. For CO, \cite{Plummer77} have determined the 
	photoionization cross section of CO$^+$(B$^2\Sigma^+$) 
	using their experimentally measured branching ratios and total absorption 
	cross section, which is also consistent with the earlier experimental  
	determination by \cite{Samson76} and \cite{Vanderwiel72}.  
\cite{Samson76} has evaluated that the statistical error in the measured 
branching ratios is  less than 10\%. Thus, the effect of uncertainty in the 
measured photoionization cross section of CO has a small impact on the modelled 
limb emission intensities.} 

{The uncertainties associated with
the measured electron impact cross sections of CO$_2$ and CO by 
\cite{Ajello71-co2} and 
\cite{Arqueros81}, respectively, are about 25\%. However, it should be noted 
that the 
photoionization excitation rates of CO$_2$ and CO producing 
CO$^+$(B$^2\Sigma^+$) are higher by a factor 2 or 
more compared to those of electron impact ionization (see 
Fig.~\ref{fig:vol_prate}). Thus, the role of 
uncertainties in the measured electron impact cross sections on the modelled 
limb intensity profiles can be neglected.} 
{Regarding the uncertainty in the solar flux, \cite{Thiemann17} have 
studied that  about a maximum of 30\%  uncertainty  is possible in the daily 
measured solar flux for the wavelengths less 
than 700 \AA\ in the  Flare Irradiance Spectral Model on Mars (FISM-M).  
Hence, 
{for given} CO and CO$_2$ density profiles,
we estimate that the 
uncertainties associated with 
the cross sections and the solar flux 
 can lead to about a maximum of 40--50\% uncertainty  in the 
 modelled 
limb intensity.}   

{On 2016 April 07, IUVS measured CO$^+$ first-negative band emission 
intensity with a standard deviation of  one $\sigma$ \citep{Stevens19}. 
But it can be noticed in Figure~\ref{fig:com_fmin} that for the altitudes above 
170 km, the observed CO$^+$ 
first-negative band emission intensity vary significantly.
Considering this variation into account, we varied the CO densities in MCD and 
MGIT models to reconcile the IUVS/MAVEN observation. We find  a factor of 
6--10 and 3--5 increment is required for CO density profiles of MGIT and MCD 
models, respectively, to explain the observed emission intensity profile. This 
calculation suggests that during IUVS observation period CO density may be 
varying 
significantly along the IUVS/MAVEN line of sight. However, there are no 
 NGIMS/MAVEN in-situ measured CO density profiles available to study  the 
variation in IUVS/MAVEN observed CO$^+$ first-negative band intensity.}

The CO is an important minor species in the Martian upper atmosphere, 
which plays a significant role in maintaining the stability of CO$_2$ via 
a 
photochemical cycle. The direct recombination of CO and O, which are 
dissociative products of CO$_2$, is forbidden. Hence, this reaction 
can not be an effective mechanism to recycle 
CO$_2$ 	 in the Martian upper atmosphere.  
By invoking heterogeneous chemistry, several 
modelling 
works were carried out to study the stability of CO$_2$ in the Martian 
atmosphere  \citep{McElroy70, McElroy76, Nier77, Krasnopolsky93, Nair94, 
	Atreya94, Atreya95}.  Since CO primarily originates from the 
photodissociation of CO$_2$, the variation in CO$_2$ density will have a 
direct 
impact on the CO abundance. Besides the variation in CO$_2$ neutral 
density, 
the  chemical loss reaction between OH and CO, which 
consequently recycle the CO$_2$, also determines the 
concentration of CO in the Martian upper atmosphere 
\citep{Nair94,Gonzalez05}.
As shown earlier, when we 
use a larger amount of CO  than MCD and MGIT modelled values, the 
modelled 
limb 
intensity profile for case-A and case-B  are in agreement  with the 
IUVS/MAVEN 
observation (see Fig.~\ref{fig:com_fmin}). 
Since  no strong variability in the NGIMS/MAVEN measured CO$_2$ 
number density is seen (with a variation of factor 2 from the mean 
value, see 
Fig.~\ref{fig:numb_neu}) and also the derived thermosphere temperature 
is low
\citep[$\sim$200 K, see][]{Bougher15}, the larger amount of CO density in 
the 
Martian upper atmosphere is only possible {if} its chemical loss rate is 
significantly low  during the IUVS observation period. By analysing the 
Compact 
Reconnaissance Imaging Spectrometer for Mars (CRISM) instrument onboard 
Mars 
Reconnaissance Orbiter (MRO) data, \cite{Smith18} have 
also shown that CO 
density can undergo a strong seasonal variation. Hence, the collisional 
chemistry of CO 
strongly determines its  neutral density in the Martian upper atmosphere 
and consequently the observed CO$^+$ first-negative emission intensity.

{As shown in Figure~\ref{fig:numb_neu}, the  CO$_2$ density profiles of 
MCD and MGIT models are closer to the NGIMS/MAVEN in situ 
measurements. However,  it should be noticed 
that the CO$_2$ density of  MCD 
model is 
smaller than that of MGIT model by a factor 3 to 5. 
But in contrast, the CO density  of MCD model is higher 
compared to that of MGIT model by a factor 2 to 10. The
differences between CO and CO$_2$ densities of these models lead to 
an order of 
magnitude higher CO/CO$_2$ volume mixing ratio in the MCD model when compared 
to 
that in the MGIT model (see  
dashed curves in Fig.~\ref{fig:numb_neu}). 
This significant change in the CO/CO$_2$ {volume mixing} ratios} may be 
due to the  photochemistry of CO incorporated in 
these general circulation models. Though the major photochemical
production and loss reactions of CO are incorporated in these models, 
{the CO density profiles in both models} are not sufficient to explain 
the IUVS/MAVEN 
observation. This suggests that probably the CO density profiles are 
underestimated during the IUVS/MAVEN observation period. 
Hence, based on our model calculations, we suggest that the role of CO is 
more important than that of CO$_2$ and the observed limb 
intensity profile of this band emission can be 
used as a good tracer to study the CO {volume} mixing ratio and also 
its seasonal variation for the altitudes above 150 km.  
The retrieval of CO density profiles based on the CO$^+$ 
first-negative band also helps to understand the production and loss of 
CO, which is primarily linked with CO$_2$ density and also collisional 
chemistry, in the Martian upper atmosphere. {However, it should be noted 
the thermospheric temperature can play an important role in  determining the 
CO$_2$ density at the altitudes above 150 km.  
 Using CO$_2^+$ ultraviolet doublet 
emission at 289 nm, \cite{Jain20} recently measured first-ever 
 thermospheric temperature at an altitude of 170 km during the planetary 
 encircling dust 
event of 2018. They observed that the mean temperature of 
thermosphere can be  increased as high as 20 K at higher latitudes during the 
global dust event.} 
Thus for high thermosphere temperature 
(about 300 K), the contribution of CO$_2$, via photon and electron 	
impact  dissociative ionization, can also be significant to the total 
CO$^+$ first-negative band emission.
 Hence, modelling of this band 
emission by incorporating the primary production processes is 
essential to derive the CO density based on the observed emission 
intensity profiles.

\section{Summary and Conclusions}
\label{sec:sum_con}
IUVS onboard MAVEN mission recently observed first-ever CO$^+$  
first-negative 
band emission in the Martian upper atmosphere \citep{Stevens19}. We have 
studied the photochemistry of this band emission  by incorporating various 
CO$^+$(B$^2\Sigma^+$) formation processes and the neutral 
densities of the upper atmosphere from MCD and MGIT models. By 
comparing the 
modelled limb intensity profiles with the IUVS/MAVEN measurements, we 
found that 
CO$_2$ is not a suitable candidate to explain the observed intensity 
profile of 
CO$^+$ first-negative band emission. By increasing  the input CO density 
(by a 
factor of 3 and 8 in MCD and MGIT models, respectively), 
the 
modelled   CO$^+$  first-negative band intensity profile is in agreement 
with 
IUVS/MAVEN observation, which suggests that a large amount of CO was 
present during the observation period.	In this case, the observed 
emission intensity is 
significantly governed 	by the ionization of CO, by both photons and 
electrons.  Since CO significantly contributes to this
emission, we suggest that the observed band intensity can be used to 
retrieve 
CO abundance in the Martian upper atmosphere for the altitudes above 150 
km. The derivation of CO density 
based on the observed CO$^+$ first-negative emission intensity profile 
also
helps to 
constrain its {volume} mixing ratio in the general circulation models as 
well as to 
study the stability of CO$_2$ in the Martian upper atmosphere.
 More observations of this band 
emission along with modelling are essential to constrain the CO {volume} 
mixing 
ratio 
and its 
variation during different seasonal conditions in the Martian upper 
atmosphere. 

\begin{acknowledgements}
The neutral and ion number densities used in the present study have 
been taken from
MAVEN/NGIMS data which is accessible through the web link 
https://pds-atmospheres.nmsu.edu. We would like to thank Prof. Stephen 
Bougher for providing the neutral density profiles from MGIT model.
SR is supported by Department of Science 
and Technology (DST) with Innovation in Science Pursuit for 
Inspired Research (INSPIRE) faculty award 
[grant:dst/inspire/04/2016/002687], and he would like to 
thank Physical Research Laboratory for facilitating 
conducive research environment. The authors would like to thank the  reviewer 
for the valuable comments and suggestions that improved the manuscript. 
\end{acknowledgements}

\end{document}